\def   \ni {\noindent}

\def   \ssk {\vskip  5truept}

\def   \bsk {\vskip 15truept}
 
\def   \newpage {\vfill\eject}
\def   \newline {\hfil\break}

\def\phibar{\overline\varphi}

\documentstyle[epsfig]{article}
\begin{document}

\hsize 5truein
\vsize 8truein
\font\abstract=cmr8
\font\keywords=cmr8
\font\caption=cmr8
\font\references=cmr8
\font\text=cmr10
\font\affiliation=cmssi10
\font\author=cmss10
\font\mc=cmss8
\font\title=cmssbx10 scaled\magstep2
\font\alcit=cmti7 scaled\magstephalf
\font\alcin=cmr6 
\font\ita=cmti8
\font\mma=cmr8
\def\ref{\par\noindent\hangindent 15pt}
\null

\setlength{\unitlength}{1mm}
\def\fwb{63mm}
\def\fhb{50mm}
\newcommand{\Dunits}{$\times10^{28}$ cm$^2$s$^{-1}$}
\newcommand{\gray}{$\gamma$-ray\ }
\hyphenation{brems-strahl-ung}

{\footnotesize \it \vspace{-14\baselineskip} \noindent 
   Proc.\ 3rd INTEGRAL Workshop ``The Extreme Universe'',
   14--18 Sep.\ 1998, Taormina, Italy
   \\ \rule[3ex]{124mm}{0.1mm}
\vspace{1ex} \vspace{11\baselineskip} }


\title{\ni COMPTEL skymapping: a new approach using

\ni parallel computing.   }                                               

\bsk \bsk
\author{\ni  A.W. Strong$^1$, H. Bloemen$^2$, R. Diehl$^1$, W. Hermsen$^2$, V. Sch\"onfelder$^1$  }                                                       
\bsk
\affiliation{1) Max-Planck-Institut f\"ur extraterrestrische Physik, 
 Garching, Germany}                                                

\affiliation{2) SRON-Utrecht, Utrecht, The Netherlands}                                            
\bsk
\baselineskip = 12pt

\abstract{ABSTRACT \ni
Large-scale skymapping with COMPTEL using the full
survey database presents challenging problems on
account of the complex response and time-variable
background.
A new approach which attempts to address some of these
problems is described, in which the information about
each observation is preserved throughout the analysis.
In this method,  a maximum-entropy algorithm is used to 
determine image and background simultaneously.
Because of the extreme computing requirements, the
method has been implemented on a parallel computer,
which brings large gains since the response computation
is fully parallelizable. The zero level is left undetermined in this method.
Results using data from 7 years of COMPTEL
data are  presented.

}                                                    
\bsk
\baselineskip = 12pt

\bsk
\baselineskip = 12pt


\text{\ni 1. INTRODUCTION
\ssk
\ni     

The goal of this work is to produce quantitatively correct intensity
  maps which are free from astrophysical biases, in order to explore
  unknown emission in the MeV regime from COMPTEL.  The high instrumental background level of COMPTEL implies that imaging methods require a good estimate of the background. In the absence of an independent (`off') measurement
  for our case of  continuum  (as opposed to spectral-line) emission, 
one approach is to use high-latitude observations, where essentially only the instrumental and cosmic diffuse backgrounds are present, as a template for the background model. 
The assumption that the background in the instrumental system has a constant form in at least the spatial coordinates (but not in Compton scatter angle) may  serve as the basis for an imaging method. The idea is to construct skymaps using all observations simultaneously and explicitly,
preserving the instrument-system for the dataspace and preserving the specific information for each observation.
 Since COMPTEL has more than 240 observations
covering the whole sky this procedure requires very large CPU resources. 
 This is the reason why such an approach had to await the
  feasibility of parallel computing
\bsk
\ni 2. METHOD 
\ssk
\ni 
It is assumed that the COMPTEL response for one observation $i$ to an intensity distribution  $I_\gamma$  has the usual form in terms of  the expected counts $n_i$ :
$$n_i(\chi,\psi,\phibar)=g_i(\chi,\psi,\phibar)\int\int  I_\gamma(\chi^\prime,\psi^\prime)X_i(\chi^\prime,\psi^\prime) f(\chi^\prime,\psi^\prime,\chi,\psi,\phibar)d\Omega + B_i(\chi,\psi,\phibar)$$ 
\newpage
\noindent where $\chi,\psi$ are the spherical coordinates of the scattered photon direction,
 $\phibar$ is the 
Compton scatter angle, 
  $g_i$ are the geometrical factors , $X_i$ is the exposure  and f is the point-spread function, $\Omega$ is the solid angle. $ B_i(\chi,\psi,\phibar)$ is the instrumental background of the observation.
The  dataspace of observations consists of  all the $n_i$ and  the response is all the $X_i$ and $g_i$.
The background is represented as 
$ B_i(\chi,\psi,\phibar) = B_o(\chi,\psi,\phibar)b_i(\phibar)$
where $B_o$ is a template derived for example from many high-latitude observations, and $b_i$ is the scaling factor for the observation $i$ for each $\phibar$. The $b_i$ are treated as free parameters to be determined, so that finally we only use the ($\chi,\psi$)-distribution of the template. The $b_i$ thus
include both temporal and $\phibar$ variations of the background, and
give n$_{\phibar}$ parameters per observation.
However for cases where the $\phibar$-distribution is known to be stable (for example above 10 MeV) it can optionally be taken from the template, so that
only temporal variations are considered, with only one parameter per observation.
The time-dependence of the background is a by-product of the method which can be compared with
other determinations.
Using high latitude observations the template is generated using
$B_o = {\Sigma n_i(\chi,\psi,\phibar)\over\Sigma g_i(\chi,\psi,\phibar)\overline{X_i}}$.
Since  $n_i$ and $g_i$  are always in the instrument system they are  aligned for all observations, so that no interpolation is required.
 The present imaging method is  differential so that  the  zero level of the maps is undetermined.


 \begin{figure}
\begin{picture}(120, 50)(20,0)
\put( 20,  0)
{\makebox(0,0)[lb] {\psfig{file=
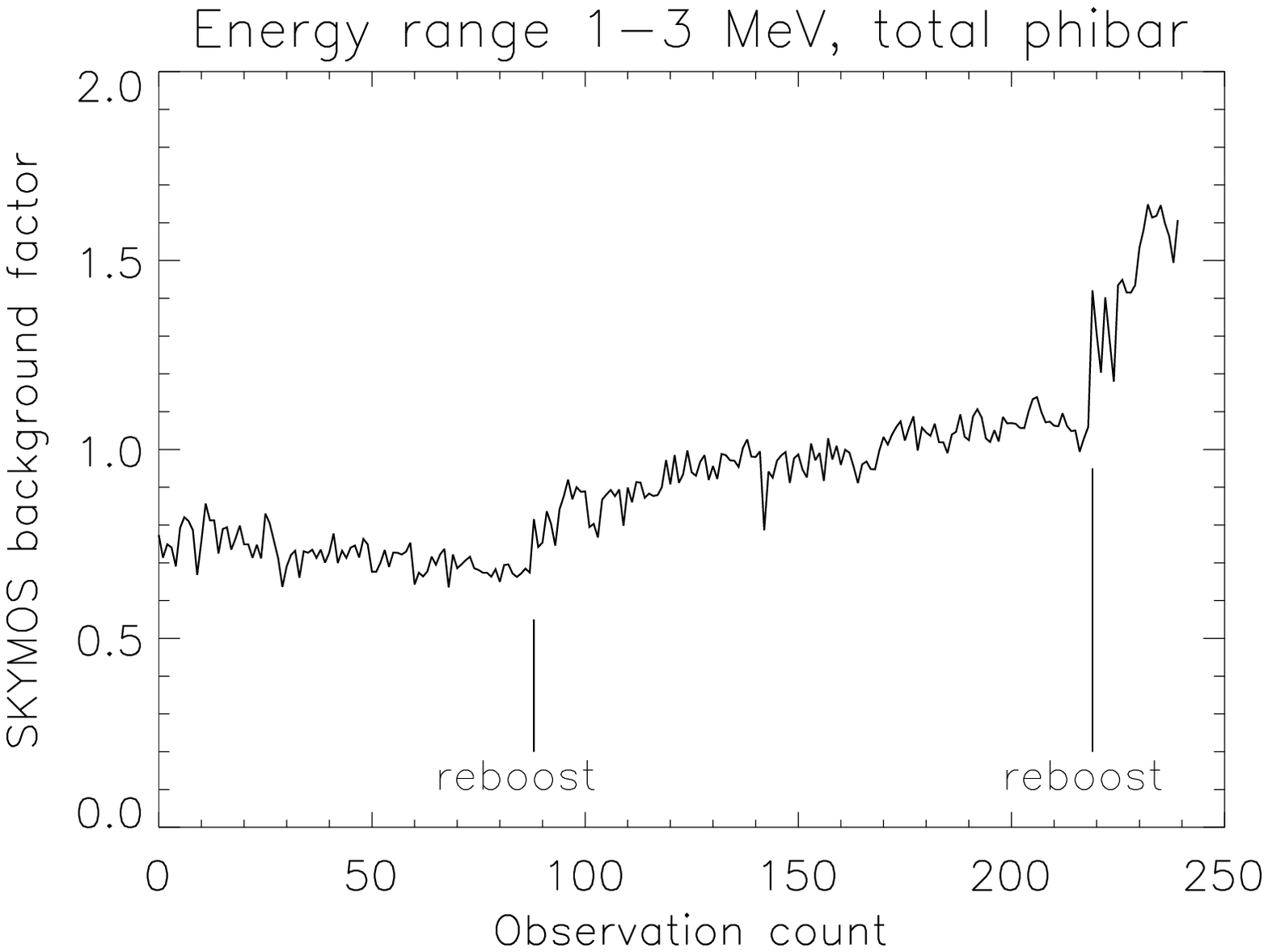,height=4.5cm}}}

\put(80,  0)
{\makebox(0,0)[lb] {\psfig{file=
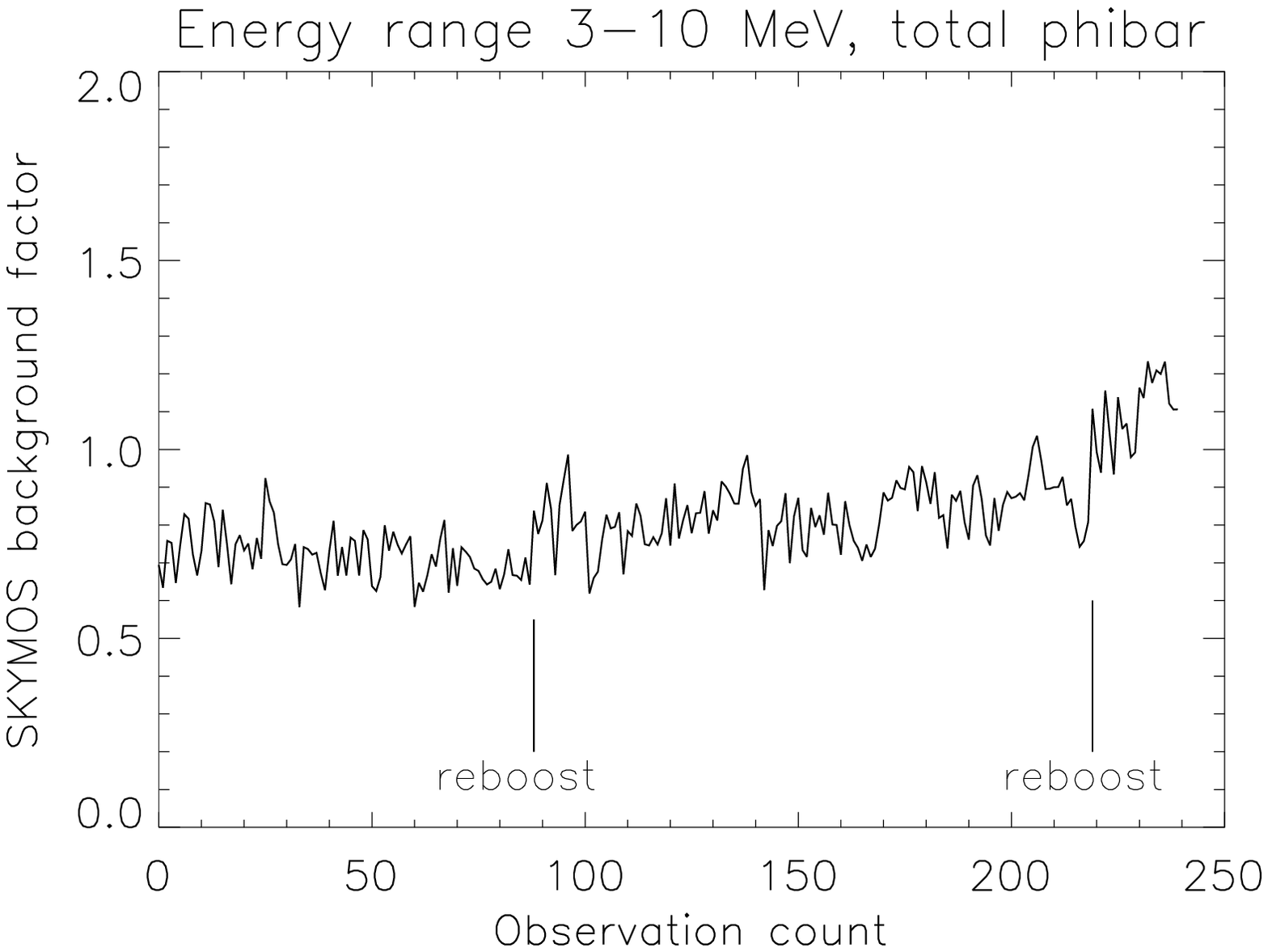,height=4.5cm}}}


\end{picture}

\caption{FIGURE 1. Relative background estimate as function of time for 1-3 and  3-10  MeV. }
\end{figure}
The parallelizing technique is to assign one PE (processing element) to each observation,
and to run the code simultaneously on all PEs.
 Upon reaching the convolution part of the calculation
each PE computes only the response for its assigned observation; the result is then transmitted to all other PEs so that each obtains the full response as required for the imaging.
In this way the `acceleration factor' is practically equal to the number of PEs.
For full-sky images one convolution requires about 5 minutes CPU on a Cray T3E, and one  iteration
requires about 10 convolutions. This gives about 1 iteration per CPU hour.
 Since the aim is to obtain intensity maps which can be used quantitatively, the iterations were
continued until saturation in intensity was reached. The effect of this is also to
produce over-structured maps on small scales so that a smoothing was applied before plotting.

\pagebreak

\begin{picture}(120,240)(-20,-20)
 \rotatebox{90}{
\put(20, 97)
{\makebox(0,0)[lb] {\psfig{file=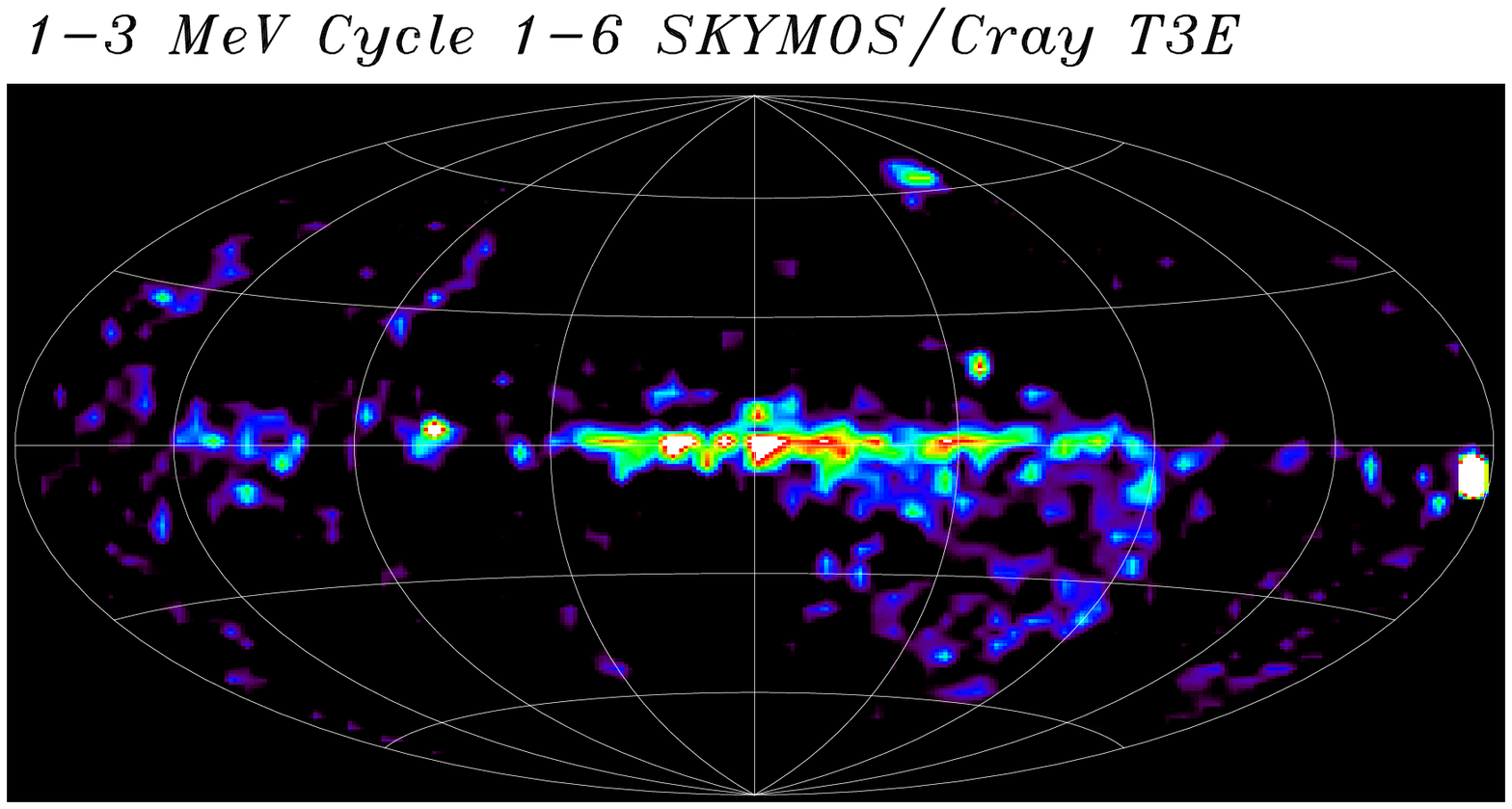,height=5cm}}}
\put(20, 44)
{\makebox(0,0)[lb] {\psfig{file=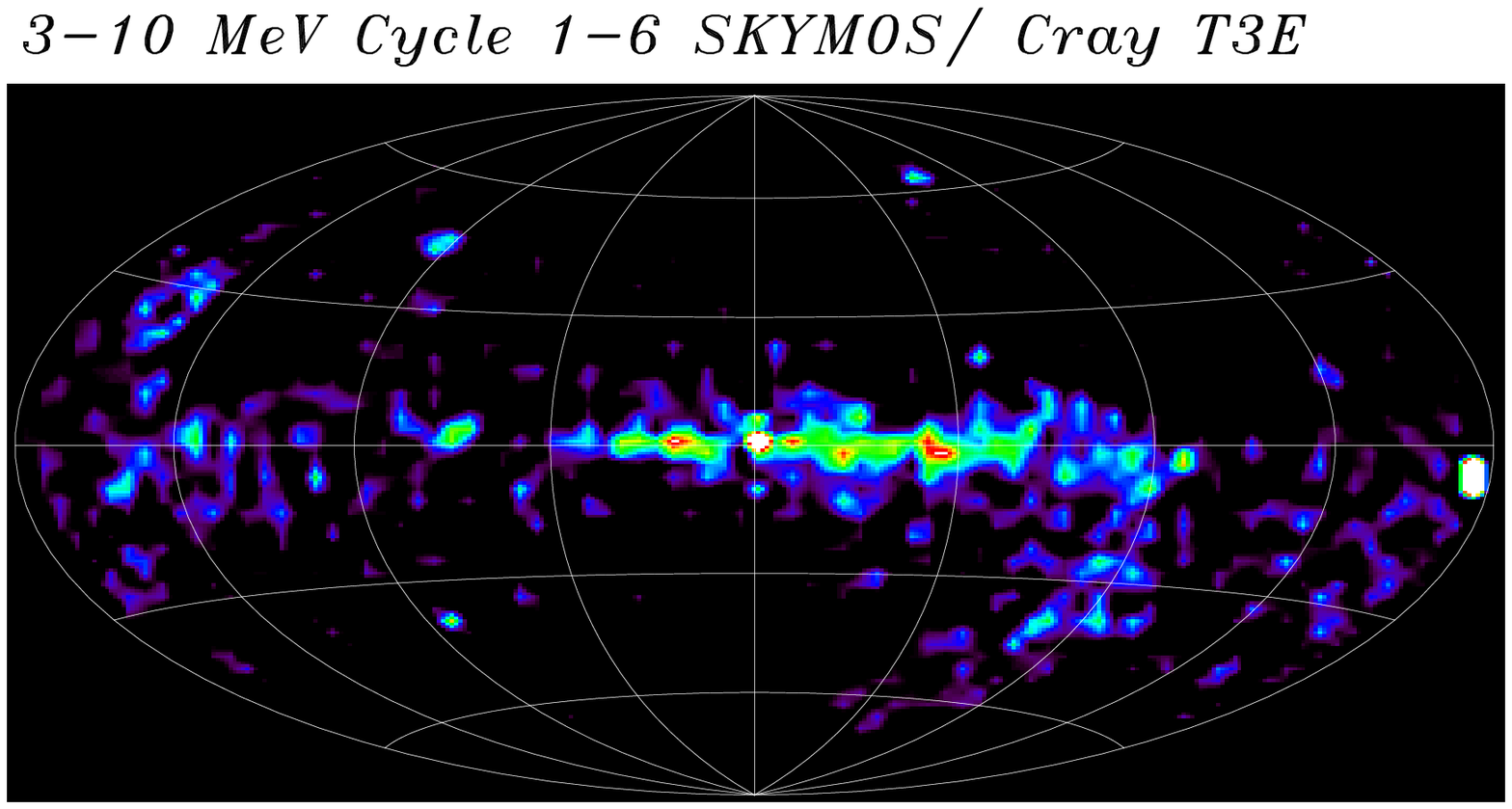,height=5cm}}}
\put(20,-9 )
{\makebox(0,0)[lb] {\psfig{file=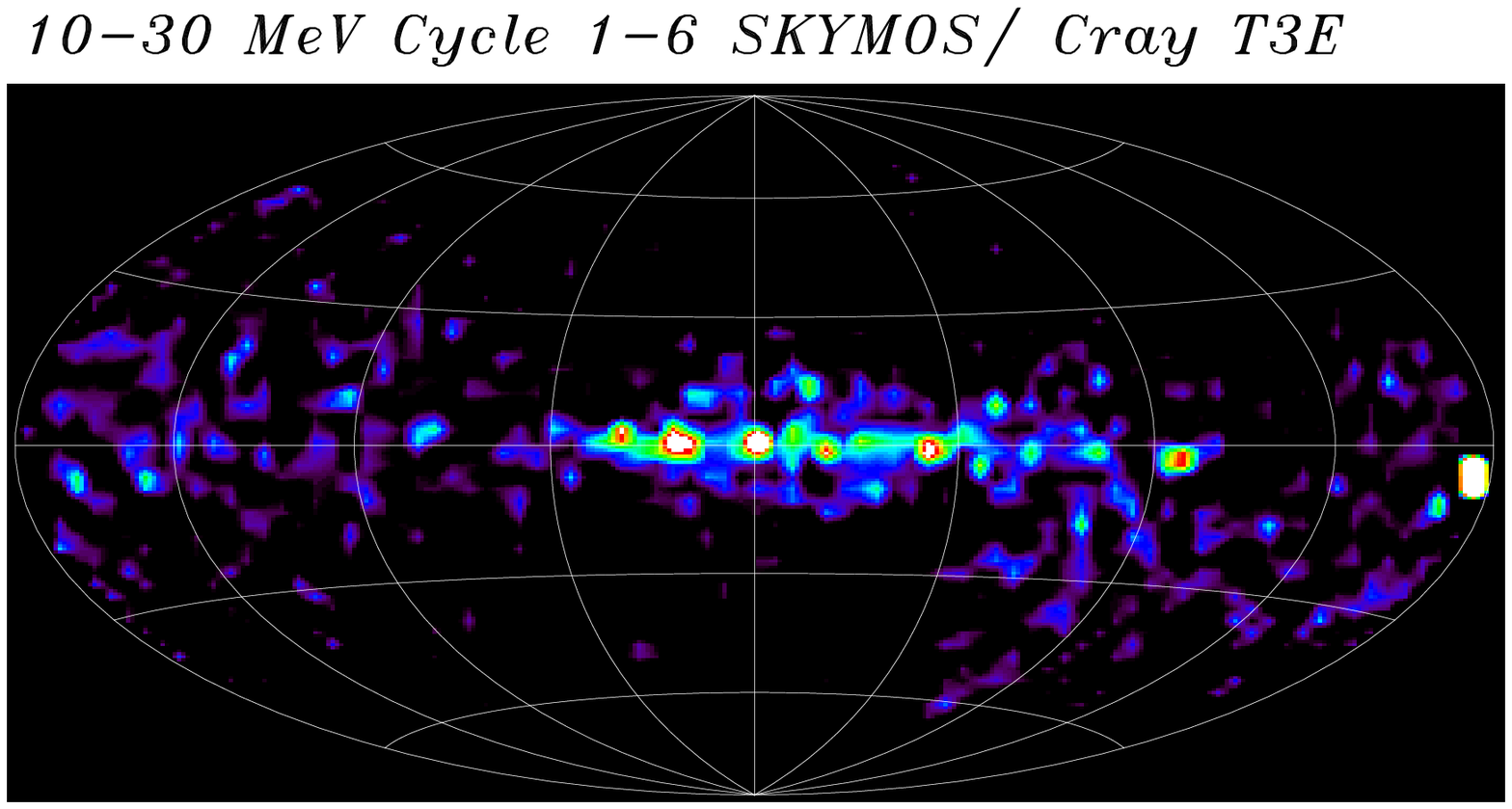,height=5cm}}}
\put(120, 98)
{\makebox( 0, 0)[lb]{\psfig{file=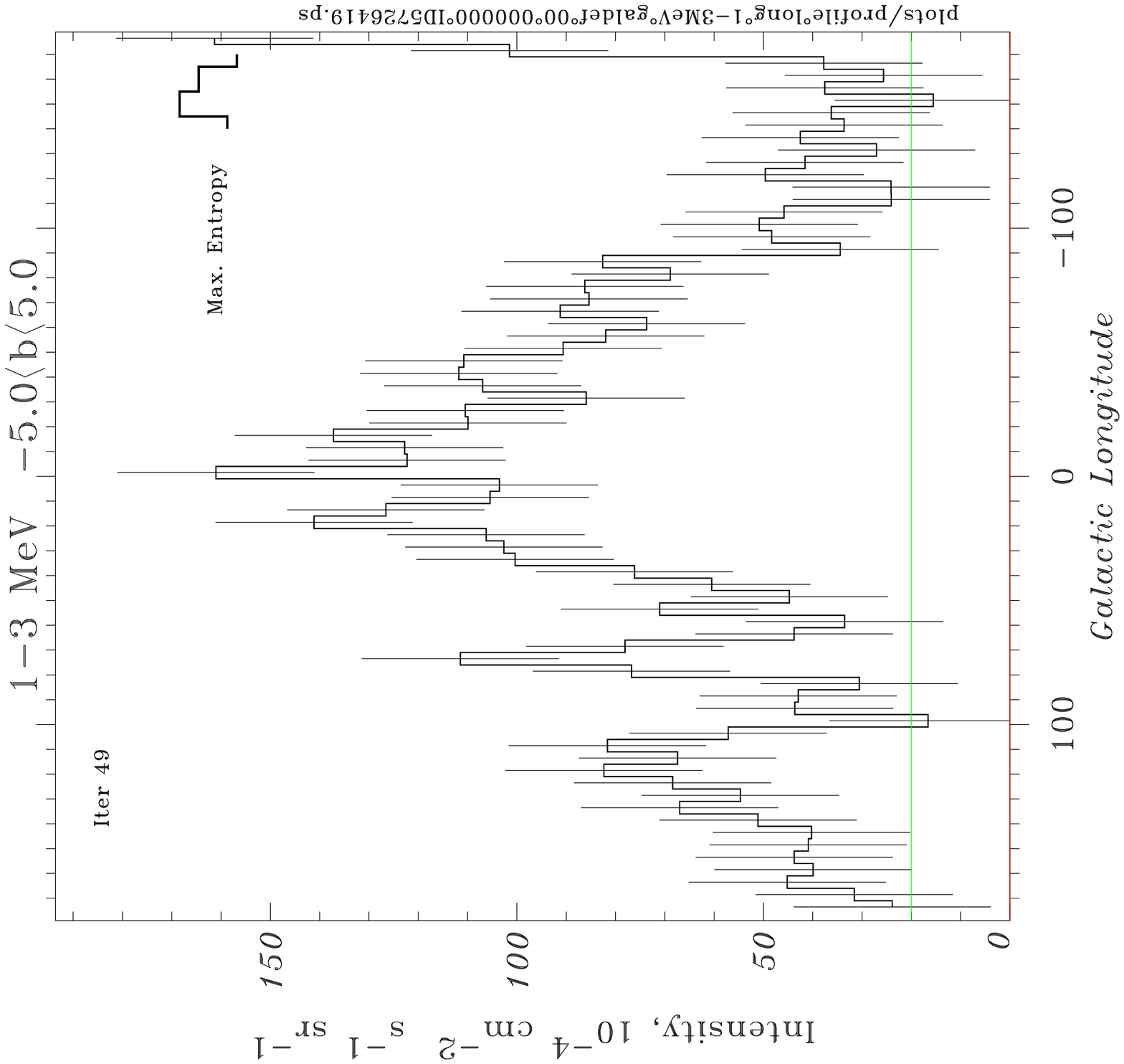,height=5cm}}}
\put(170, 98) 
{\makebox(0,0)[lb]{\psfig{file= 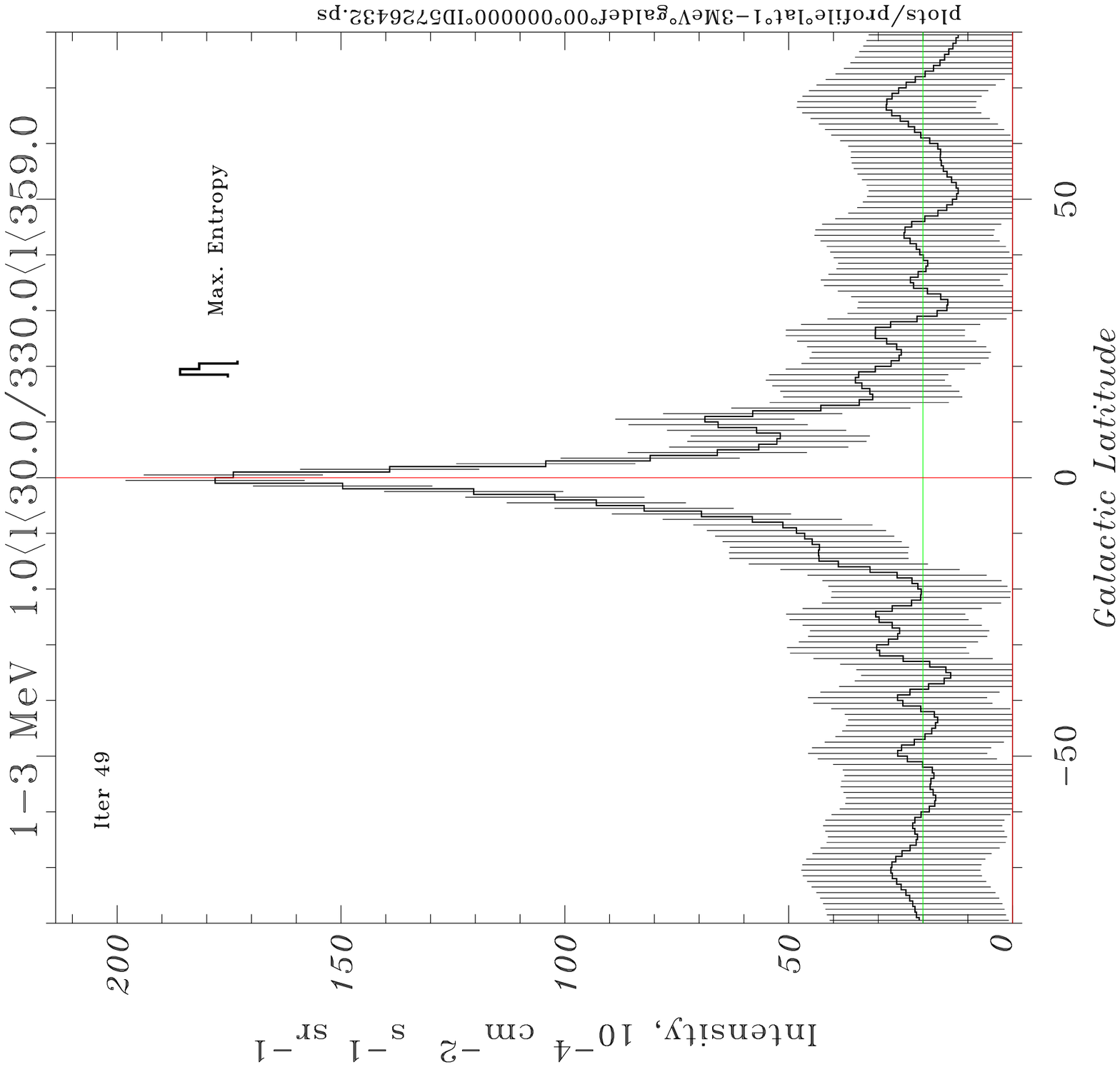,height=5cm}}}
\put(120,43)
{\makebox(0,0)[lb]{\psfig{file= 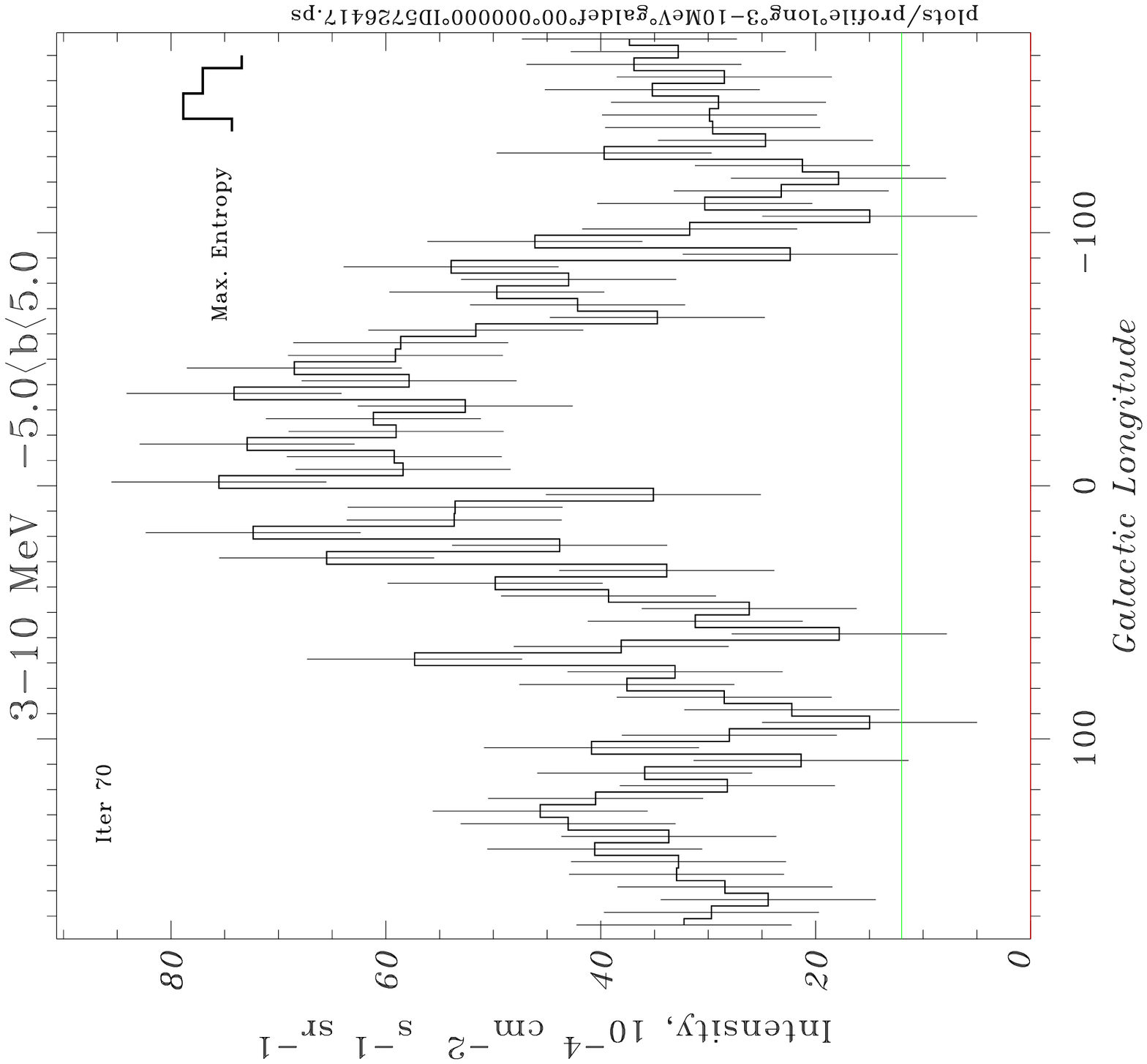,height=5cm}}}
\put(170, 43)
{\makebox(0,0)[lb]{\psfig{file= 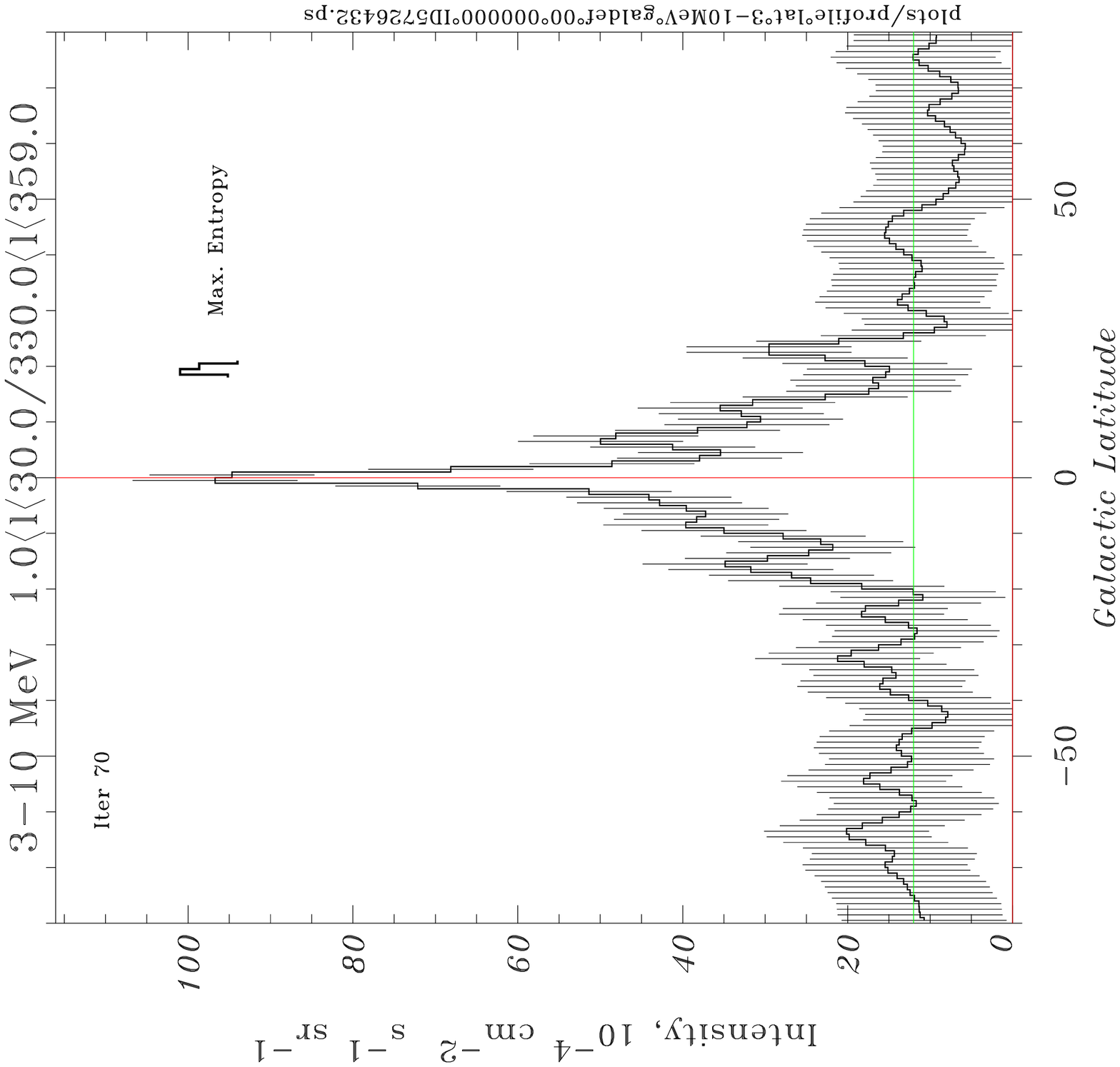,height=5cm}}}
\put(120,-10)
{\makebox(0,0)[lb]{\psfig{file= 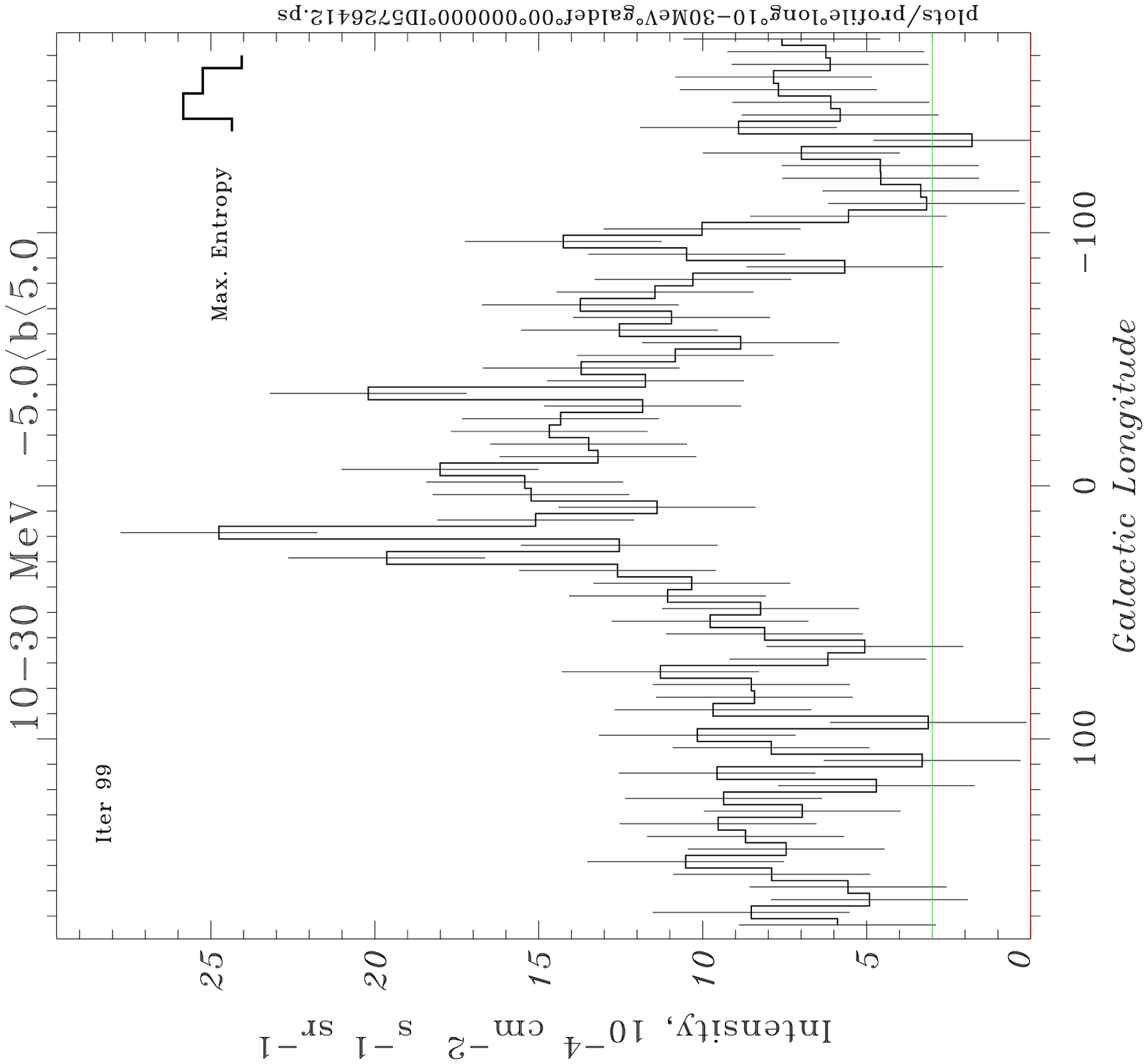,height=5cm}}}
\put(170,-10)
{\makebox(0,0)[lb]{\psfig{file= 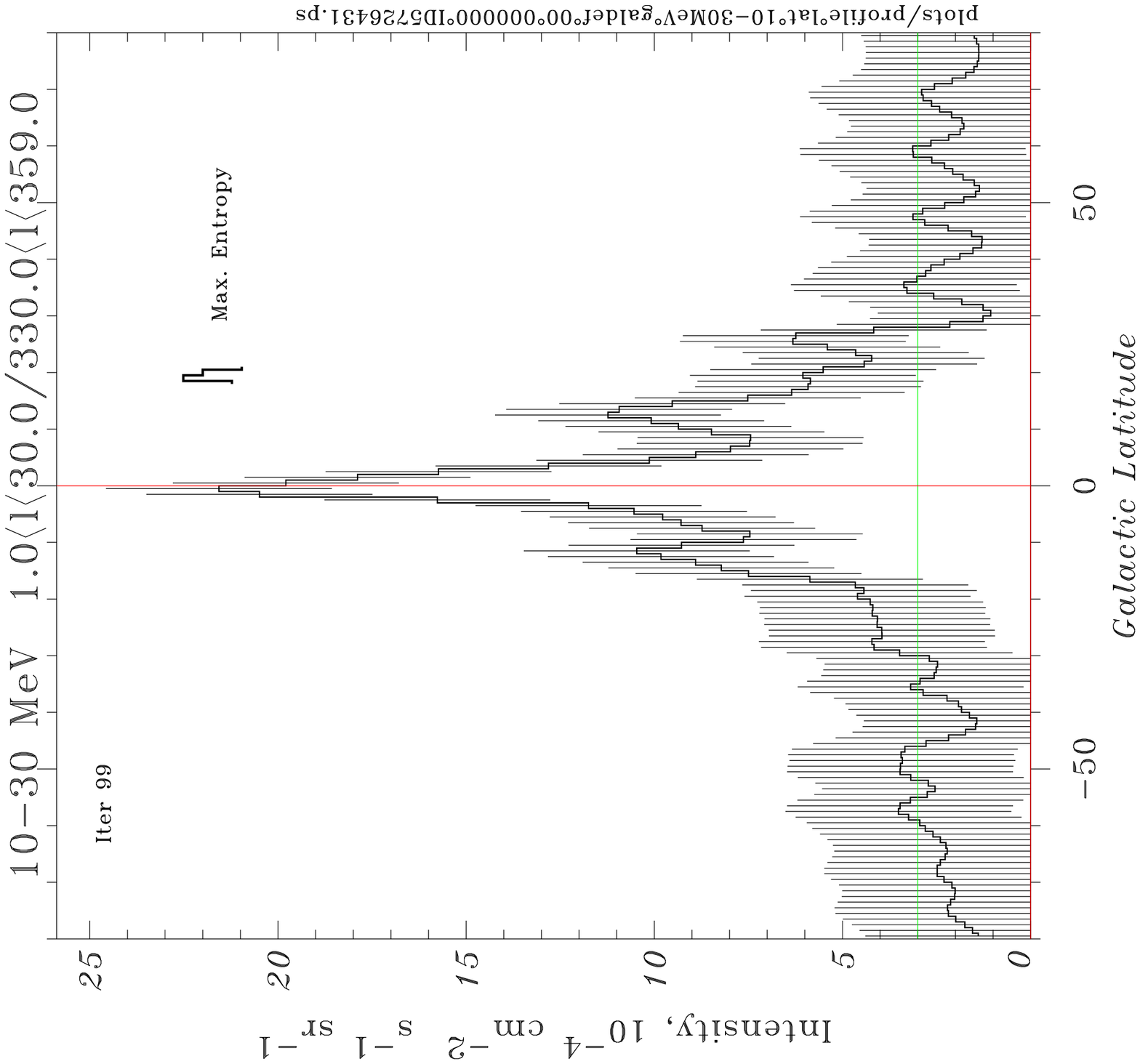,height=5cm}}}
\put(25,-20){FIGURE 2: Full sky intensity maps and longitude, latitude profiles for 1-3, 3-10 and 10-30 MeV.}
\put(45,-25){Green lines in profiles indicate {\it approximate} base-level.}                               
}

\end{picture}

 3. RESULTS
\ssk
\ni

The method was applied to all 240 observations from Cycles 1-6, for 1-3, 3-10 and 10-30 MeV.        
Fig 1 shows the derived variation of background with time, relative to the average high-latitude value.
The increase due to reboosts is visible, as expected.
The temporal behaviour agrees with detailed activation background
   histories derived from spectral analysis of identified background
   features (Varendorff et al. 1997).
Fig 2 shows skymaps and longitude, latitude profiles. 
The  skymaps have been smoothed  over 5$^o$ wide bins.
The longitude profiles are for $|b|< 5^o$ with 5$^o$ longitude bins, the latitude profiles are for $330^o<l<30^o$
with  1$^o$ latitude bins. The error bars are estimated from the fluctuations in the maps at high latitudes where
little structure is expected. The zero level of the sky intensity is not determined in this method, so these plots
have an arbitrary baseline. A representative high-latitude level is indicated, but note that this is not unique.

\begin{picture}(60,60)(0,0)
\put(-5,-5){
 \psfig{file= 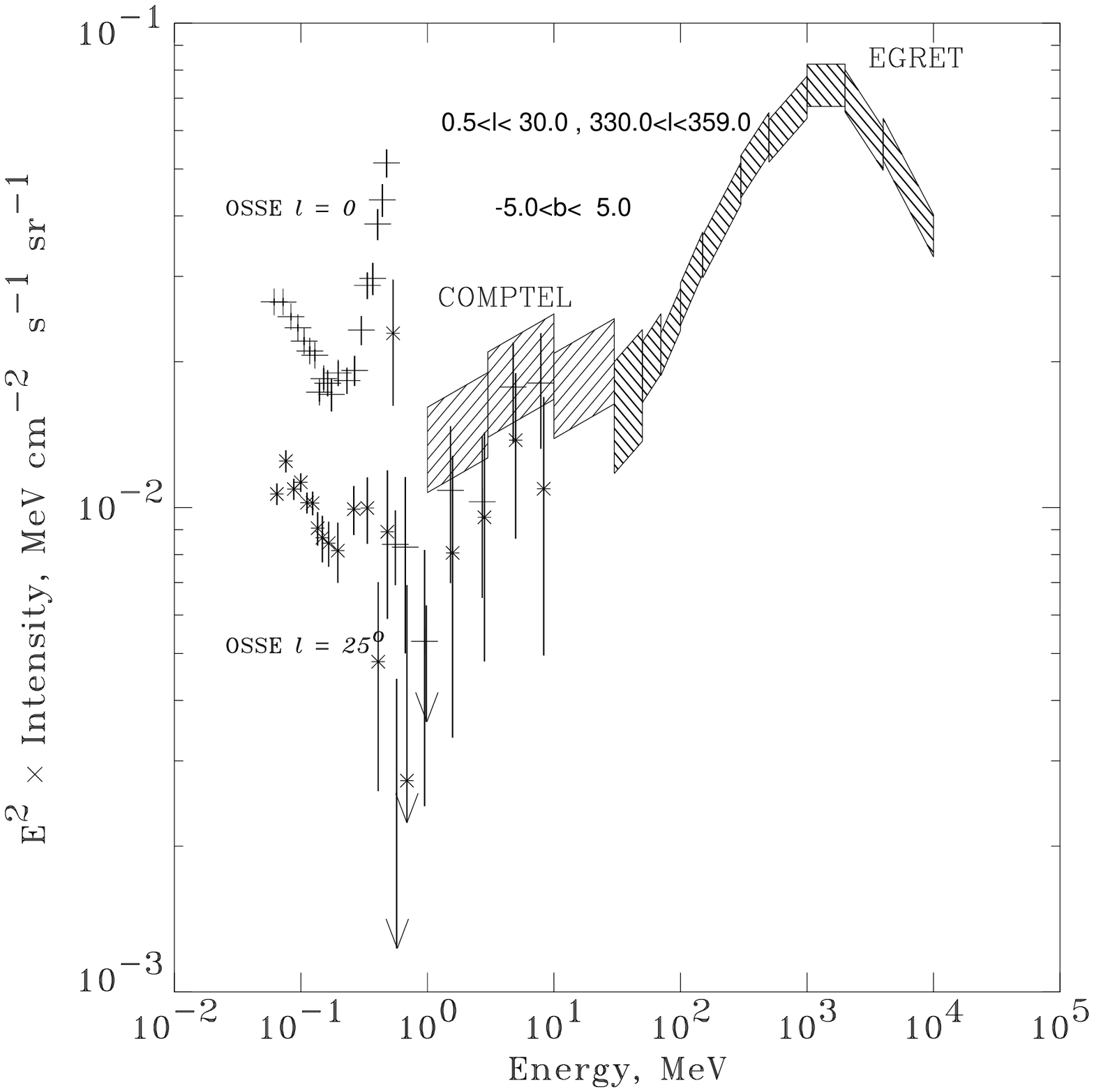,height= 6.5cm}
}
\put(60,45){FIGURE 3.}
\put(60,40){Spectra of inner Galactic plane }
\put(60,35){ $330^o<l<30^o$ $|b|<5^o$.}
\put(60,28){COMPTEL:  this work}
\put(60,21){EGRET spectrum based on }
\put(73,16){   Strong and Mattox (1996)}
\put(60,10){OSSE spectra from Kinzer et al. (1997)}
\end{picture}

\bsk
Figure 3 shows the spectrum of the inner Galactic plane for $|b|<5^o$ based on these maps, using the
high-latitude intensity as baseline.
 It is higher by about a factor 2 than given in Strong et al. (1997), 
which was our first attempt to use continuum maximum-entropy maps quantitatively 
 to derive intensities.
The explicit background treatment in the present approach represents an
advance over the earlier approach.
Recent application of an alternative technique is described in Bloemen et al. (1998).


\bsk
\baselineskip = 12pt


{\references \ni REFERENCES
\ssk

\ref Bloemen, H.   et al.\ 1998, these proceedings, paper \#160
\ref Kinzer, R. L. et al.   1997,   4th Compton Symp.\ AIP 410, New York, p. 1193; OSSE Preprint \#89.
\ref Varendorff, M. et al.   1997,   4th Compton Symp.\ AIP 410, New York, p. 1577
\ref Strong, A.W., Mattox, J.R.\ 1996, A\&A, 308, L21
\ref Strong, A.W.  et al.\ 1997,  4th Compton Symp.\ AIP 410, New York, p. 1198

}

\end{document}